# Probable cluster decays from $^{270-318}$118 superheavy nuclei

K. P. Santhosh* and B. Priyanka

*School of Pure and Applied Physics, Kannur University, Swami Anandatheertha Campus, Payyanur 670 327, Kerala, India*

email: drkpsanthosh@gmail.com

**Abstract.**

The cluster decay process in $^{270-318}$118 superheavy nuclei has been studied extensively within the Coulomb and proximity potential model (CPPM), thereby investigating the probable cluster decays from the various isotopes of Z = 118. On comparing the predicted decay half lives with the values evaluated using the Universal formula for cluster decay (UNIV) of Poenaru et al., the Universal Decay Law (UDL) of Qi et al., and the Scaling Law of Horoi et al., it was seen that, our values matches well with these theoretical values. A comparison of the predicted alpha decay half life of the experimentally synthesised superheavy isotope $^{294}$118 with its corresponding experimental value shows that, our theoretical value is in good agreement with the experimental value. The plots for $\log_{10}(T_{1/2})$ against the neutron number of the daughter in the corresponding decay reveals the behaviour of the cluster half lives with the neutron number of the daughter nuclei and for most of the decays, the half life was found to be the minimum for the decay leading to a daughter with N = 184. Most of the predicted half lives are well within the present experimental upper limit ($10^{30}$s) and lower limit ($10^{-6}$s) for measurements and hence these predictions may be of great use for further experimental investigation on cluster decay in the superheavy region.

## 1. Introduction

The mysteries behind the forces that bind the nucleonic as well as sub-nucleonic matter have been unraveled by the nuclear scientists ever since the epoch making discovery of radioactivity in 1896 [1]. Since then, several interesting experimental as well as theoretical studies have been done on the common modes of nuclear decay, such as, alpha, beta and gamma decay in the early part of the twentieth century and are now been followed by the synthesis of nuclei away from the line of stability. The discovery of cluster radioactivity, double beta decay, one and two proton radioactivity and beta delayed particle emission – the new exotic decay modes of nuclei, are the outcomes of these studies. The hype on the existence and emission of α-like clusters and clusters heavier than α particle, from both heavy and superheavy nuclei has received much attention, particularly in recent times [2-10]. Despite of so many studies, the question on the probability for the emission of heavy clusters from superheavy nuclei still remains unsolved.

The binary radioactive decay into two nuclear fragments $A_1$ and $A_2$, for which $Q > 0$ can be observed from all the nuclei with $Z > 40$; where the case $A_2 = 4$ corresponds to alpha decay, the cases in which $A_1 = A_2$ correspond to spontaneous fission and the intermediate case $4 < A_2 < A_1$ correspond to heavy-particle radioactivity. Even though thousands of two-body decays with positive $Q$ values are possible from heavy nuclei, the decay rates are undetectably low for most combinations of daughter and cluster. But, those decays with combinations corresponding to fragments with nearly closed shells may be detectable and in most cases, those decay modes having high enough $Q$ value with the heavier fragment close to $^{208}$Pb (with large mass defect due to its closed neutron and proton shells) and a tightly bound, even-even, neutron-rich lighter fragment, usually have detectable decay rates. The very low branching ratios of cluster decay ($\approx 10^{-12}$ relative to alpha decay) explains the reason for heavy particle radioactivity to be remained undiscovered until many decades after alpha radioactivity and spontaneous fission [11].

The quantum re-arrangement of large number of nucleons from the ground state of a nucleon system to two ground states of the cluster and the daughter nucleus systems was discovered by Rose and Jones at the Oxford University [12] in 1984, who measured $^{14}$C emission from $^{223}$Ra, confirming the theoretical prediction made in 1980 by Sandulescu et al., [13]. The experiment was further confirmed independently by Aleksandrov et al [14]. During the later years, the spontaneous emission of a heavier cluster, namely $^{24}$Ne from $^{231}$Pa, $^{233}$U and $^{230}$Th, was detected by Sandulescu et al., in Dubna [15] and these results were very soon re-confirmed by Price and co-workers in Berkeley [16]. The years after these major developments perceived tremendous improvements in the experimental techniques which lead to the detection of the emission of a number of clusters from $^{14}$C to $^{34}$Si from more than two dozen nuclei ranging from $^{221}$Fr to $^{242}$Cm [17, 18] and up till now, the largest known half-life in heavy cluster radioactivity is $10^{29}$ seconds for the neon emission. Even if the radioactive decay by the emission of fragments heavier than α particle is usually referred to as cluster radioactivity, the cold fission of heavy nuclei and the cold heavy cluster formation in quasi-fission may also be identified along with cluster radioactivity due to certain characteristics of these processes [19].

The various scrutinised studies done by Sandulescu et al., [20-24] on the cold valleys for binary radioactive fission has lead to yet another important discovery, the production of superheavy nuclei [25]. The core idea of these studies was to overcome the phenomenon of quasifission which interdicts the possibility to obtain superheavy elements by using the cold fragmentation valleys in the potential energy surface between different combinations giving the same compound nucleus and further it was shown that the most favorable combinations with $Z \geq 104$ are connected with Pb potential valley, the same valley of the heavy cluster emission [26]. This paved the way for the synthesis of superheavy nuclei with $Z \leq 112$ via the cold fusion reactions [27,28] in Darmstadt and the heaviest element known so far is $Z = 118$ [29, 30].

The phenomena of cold fusion, cold fission and the heavy particle radioactivity can be described using the quantum mechanical fragmentation theory (QMFT) [31], from a unified point of view. The shell closure of one/both the reaction partners for fusion or that of the decay products for fission and heavy particle decay has been the unifying aspect of this theory. A quantitative model for calculating the decay rates, penetrability and the half lives of cluster decays is inevitable in order to choose the favourable cases for experimental studies. The search for a unified description of α decay, heavy particle decay and cold fission has resulted in the proposal of several theoretical models, based on the concept of quantum tunnelling through a potential barrier, by various theoretical groups. These models can be classified as the (a) cluster/α-like models and the (b) unified/fission models. The cluster models [32-35] depict a cluster of nucleons as being preformed inside the parent nucleus with a probability that decreases with cluster size, and then tunnels out without change of size or shape. In unified models [6-9, 12, 20-24, 26, 36-41], the alpha decay, heavy particle decay and the spontaneous fission are all studied under the same footing and are treated as equivalent processes, which differs only in the degree of mass asymmetry, in which the parent nucleus deforms continuously as it penetrates the nuclear barrier and reaches the scission configuration after running down the Coulomb barrier into two fragments.

The Coulomb and proximity potential model (CPPM) proposed by Santhosh et al., [39, 40] comes under the class of fission models and has been widely used for the studies [42-48] on alpha decay, cluster decay and spontaneous fission from both heavy and superheavy nuclei and also for the fusion studies [49, 50]. The major purpose in view of the experimental studies on the alpha and cluster decay processes in the superheavy region is the prediction of the doubly magic nucleus next to $^{208}$Pb and thereby indicate towards the existence of a "magic island" or the "Island of Stabilty", around Z = 120, 124 or 126 and N = 184 [51]. Hence, the theoretical studies on alpha decay and cluster decay of superheavy nuclei may always provide new paths for the future experiments. The first attempt for the synthesis of Z = 118 was done by Oganessian et al., [52] and the synthesis of $^{294}$118 was reported as the product of the 3n-evaporation channel of the $^{249}$Cf ($^{48}$Ca, xn)$^{297-x}$118 reaction, and recently, they have been successful in synthesising $^{294}$118 via the fusion of $^{249}$Cf and $^{48}$Ca [29, 30]. Recently, within the Coulomb and proximity potential model for deformed nuclei (CPPMDN), we have done an investigation on the alpha decay properties of the isotopes of the superheavy nuclei with Z = 118, within the range $271 \leq A \leq 310$ [47] and it was found that our predicted half lives of $^{294}$118 matches well with the experimental half lives. In the present manuscript, as an extension of our earlier work, we have attempted an extensive study on the cluster decay from $^{270-318}$118 superheavy nuclei, thereby probing on the feasible cluster decays from various isotopes of Z = 118.

The detail description of the Coulomb and Proximity Potential Model (CPPM) is given in Section 2. In Section 3, we have given the results and discussions on the cluster decay of the nuclei under study and the conclusion on the entire work is given in Section 4.

## 2. The Coulomb and Proximity Potential Model (CPPM)

For the touching configuration and for the separated fragments, the potential energy barrier in Coulomb and proximity potential model (CPPM) is taken as the sum of Coulomb potential, proximity potential and centrifugal potential. The simple power law interpolation as done by Shi and Swiatecki [53] is used for the pre-scission (overlap) region. Shi and Swiatecki [53] were the first to use the proximity potential in an empirical manner and later on, Malik et al., [34] have quite extensively used it in the preformed cluster model (PCM), based on pocket formula of Blocki et al., [54] given as:

$$\Phi(\varepsilon) = -\left(\frac{1}{2}\right)(\varepsilon - 2.54)^2 - 0.0852(\varepsilon - 2.54)^3 \text{, for } \varepsilon \leq 1.2511 \quad (1)$$

$$\Phi(\varepsilon) = -3.437 \exp\left(\frac{-\varepsilon}{0.75}\right), \text{ for } \varepsilon \geq 1.2511 \quad (2)$$

where $\Phi$ is the universal proximity potential. For studying fusion cross section of different target-projectile combinations, Dutt et al., [55, 56] have used different versions of proximity potentials. Another formulation of proximity potential [57] is been used in the present model, as given by Eqs. 6 and 7, and the assault frequency $v$ is calculated for each parent-cluster combination which is associated with vibration energy. But, for even A parents and for odd A parents, Shi and Swiatecki [58] got $v$ empirically, unrealistic values as $10^{22}$ and $10^{20}$ respectively.

The interacting potential barrier for a parent nucleus exhibiting cluster decay is given by,

$$V = \frac{Z_1 Z_2 e^2}{r} + V_p(z) + \frac{\hbar^2 \ell(\ell+1)}{2\mu r^2}, \text{ for } z > 0 \quad (3)$$

Here $Z_1$ and $Z_2$ are the atomic numbers of the daughter and emitted cluster, '$z$' is the distance between the near surfaces of the fragments, '$r$' is the distance between fragment centers and is given as $r = z + C_1 + C_2$, where, $C_1$ and $C_2$ are the Süsmann central radii of fragments. The term $\ell$ represents the angular momentum, $\mu$ the reduced mass and $V_P$ is the proximity potential. The proximity potential $V_P$ is given by Blocki et al. [54] as,

$$V_p(z) = 4\pi\gamma b \left[ \frac{C_1 C_2}{(C_1 + C_2)} \right] \Phi\left(\frac{z}{b}\right), \quad (4)$$

with the nuclear surface tension coefficient,

$$\gamma = 0.9517[1 - 1.7826(N - Z)^2 / A^2] \text{ MeV/fm}^2 \quad (5)$$

where $N$, $Z$ and $A$ represent neutron, proton and mass number of parent respectively, $\Phi$ represents the universal proximity potential [57] given as

$$\Phi(\varepsilon) = -4.41 e^{-\varepsilon/0.7176}, \text{ for } \varepsilon > 1.9475 \quad (6)$$

$$\Phi(\varepsilon) = -1.7817 + 0.9270\varepsilon + 0.0169\varepsilon^2 - 0.05148\varepsilon^3, \text{ for } 0 \leq \varepsilon \leq 1.9475 \quad (7)$$

With $\varepsilon = z/b$, where the width (diffuseness) of the nuclear surface $b \approx 1$ fm and Süsmann central radii $C_i$ of fragments related to sharp radii $R_i$ as,

$$C_i = R_i - \left(\frac{b^2}{R_i}\right) \quad (8)$$

For $R_i$ we use semi empirical formula in terms of mass number $A_i$ as [54],

$$R_i = 1.28 A_i^{1/3} - 0.76 + 0.8 A_i^{-1/3} \quad (9)$$

The potential for the internal part (overlap region) of the barrier is given as,

$$V = a_0(L - L_0)^n, \text{ for } z < 0 \tag{10}$$

Here $L = z + 2C_1 + 2C_2$ and $L_0 = 2C$, the diameter of the parent nuclei. The constants $a_0$ and $n$ are determined by the smooth matching of the two potentials at the touching point.

Using one dimensional WKB approximation, the barrier penetrability P is given as,

$$P = \exp\left\{-\frac{2}{\hbar}\int_a^b \sqrt{2\mu(V-Q)}\,dz\right\} \tag{11}$$

Here the mass parameter is replaced by $\mu = mA_1A_2/A$, where '$m$' is the nucleon mass and $A_1$, $A_2$ are the mass numbers of daughter and emitted cluster respectively. The turning points "$a$" and "$b$" are determined from the equation $V(a) = V(b) = Q$. The above integral can be evaluated numerically or analytically, and the half life time is given by

$$T_{1/2} = \left(\frac{\ln 2}{\lambda}\right) = \left(\frac{\ln 2}{\nu P}\right) \tag{12}$$

where, $\nu = \left(\frac{\omega}{2\pi}\right) = \left(\frac{2E_v}{h}\right)$ represent the number of assaults on the barrier per second and $\lambda$ the decay constant. $E_v$, the empirical vibration energy is given as [59],

$$E_v = Q\left\{0.056 + 0.039\exp\left[\frac{(4-A_2)}{2.5}\right]\right\}, \quad \text{for } A_2 \geq 4 \tag{13}$$

Classically, the α particle is assumed to move back and forth in the nucleus and the usual way of determining the assault frequency is through the expression given by ν = velocity/(2R), where R is the radius of the parent nuclei. But, as the α particle has wave properties, a quantum mechanical treatment is more accurate. Thus, assuming that the α particle vibrates in a harmonic oscillator potential with a frequency ω, which depends on the vibration energy $E_v$, we can identify this frequency as the assault frequency ν given in Eqs. (12) and (13).

### 3. Results and discussions

The possibility to have a cluster decay process is related to its exotermicity, $Q > 0$. The energy released in decay transitions between the ground state energy levels of the parent nuclei and the ground state energy levels of the daughter nuclei is given as

$$Q_{gs \to gs} = \Delta M_p - (\Delta M_c + \Delta M_d) \tag{14}$$

where $\Delta M_p, \Delta M_d, \Delta M_c$ are the mass excess of the parent, daughter and cluster respectively. The $Q$ values for cluster decays have been evaluated using the experimental mass excess values of Wang et al., [60] and some of the mass excesses have been taken from Koura-Tachibana-Uno-Yamada (KTUY) [61], as those experimental mass excess were unavailable in Ref [60]. In the case of $^{270}$118, the mass excess has been taken from Ref. [62]. The aim of the present work is to study the feasibility for the emission of clusters $^4$He, $^{8,10}$Be, $^{12,14,16,18}$C, $^{16,18,20,22,24}$O, $^{22,24,26,28}$Ne, $^{26,28,30,32,34}$Mg, $^{30,32,34,36,38}$Si and $^{38,40,42,44}$S from $^{270-318}$118 superheavy nuclei, using the Coulomb and proximity potential model (CPPM). In CPPM, the external drifting potential barrier is obtained as the sum of the Coulomb potential, proximity potential and centrifugal potential for the touching configuration and for the separated fragments. In order to identify the proton emitters in $^{270-318}$118 superheavy nuclei, the one-proton and the two-proton separation energies [63] of all the isotopes under study were also evaluated using the relations given as

$$S(p) = -\Delta M(A, Z) + \Delta M(A-1, Z-1) + \Delta M_H = -Q(\gamma, p) \qquad (15)$$

$$S(2p) = -\Delta M(A, Z) + \Delta M(A-2, Z-2) + 2\Delta M_H = -Q(\gamma, 2p) \qquad (16)$$

where $S(p)$ and $S(2p)$ are the one-proton and two-proton separation energies of the nuclei respectively, $\Delta M(A, Z)$ and $\Delta M_H$ represents the mass excess of the parent and the proton respectively. The terms $\Delta M(A-1, Z-1)$ and $\Delta M(A-2, Z-2)$ represents the mass excess of the daughter nuclei produced during the one-proton and the two-proton radioactivities respectively. The terms $Q(\gamma, p)$ and $Q(\gamma, 2p)$ represents the $Q$ value for the one-proton and two-proton radioactivity respectively. The evaluation of the separation energies for all the isotopes of $Z = 118$ shows that the one-proton separation energy, $S(p)$ is negative for those isotopes within the range $270 \leq A \leq 280$ and the two-proton separation energy, $S(2p)$ is negative for those isotopes within the range $270 \leq A \leq 284$. Thus, it is evident that all those isotopes within the range $270 \leq A \leq 284$ are outside the proton drip line and thus easily decays through proton emission. Hence we have limited our computation of the heavy particle decay of $Z = 118$ for only those isotopes within $286 \leq A \leq 318$.

### 3.1 Cluster decay half lives

The cluster decay half lives for the isotopes under study have also been evaluated within the Universal Decay Law (UDL) of Qi et al., [4, 64], the Universal (UNIV) curve of Poenaru et al., [7, 65] and the Scaling Law of Horoi et al., [66]. The formalisms are discussed below.

### 3.1.1 The Universal Curve (UNIV)

The decay half lives have been explained using several simple and effective relationships, which are obtained by fitting the experimental data. The universal (UNIV) curves [67-70], derived by extending a fission theory to larger mass asymmetry should be mentioned, among them, with great importance. Based on the quantum mechanical tunnelling process [71, 72], in UNIV, the disintegration constant $\lambda$, valid in both fission-like and $\alpha$-like theories and the partial decay half life $T$ of the parent nucleus is related as,

$$\lambda = \ln 2/T = \nu S P_S \tag{17}$$

Here $\nu$, $S$ and $P_s$ are three model-dependent quantities: $\nu$ is the frequency of assaults on the barrier per second, $S$ is the pre-formation probability of the cluster at the nuclear surface (equal to the penetrability of the internal part of the barrier in a fission theory [67, 68]), and $P_s$ is the quantum penetrability of the external potential barrier.

By using the decimal logarithm,

$$\log_{10} T(s) = -\log_{10} P - \log_{10} S + [\log_{10}(\ln 2) - \log_{10} \nu] \tag{18}$$

To derive the universal formula, it was assumed that $\nu$ = constant and that $S$ depends only on the mass number of the emitted particle $A_e$ [68, 71], as the microscopic calculation of the pre-formation probability [32] of many clusters from $^8$Be to $^{46}$Ar had shown that it is dependent only upon the size of the cluster. Using a fit with experimental data for $\alpha$ decay, the corresponding numerical values [68] obtained were, $S_\alpha = 0.0143153$, $\nu = 10^{22.01}\text{s}^{-1}$. The decimal logarithm of the pre-formation factor is given as,

$$\log_{10} S = -0.598(A_e - 1) \tag{19}$$

and the additive constant for an even-even nucleus is,

$$c_{ee} = [-\log_{10} \nu + \log_{10}(\ln 2)] = -22.16917 \tag{20}$$

The penetrability of an external Coulomb barrier, having separation distance at the touching configuration $R_a = R_t = R_d + R_e$ as the first turning point and the second turning point defined by $e^2 Z_d Z_e / R_b = Q$, may be found analytically as

$$-\log_{10} P_S = 0.22873(\mu_A Z_d Z_e R_b)^{1/2} \times [\arccos\sqrt{r} - \sqrt{r(1-r)}] \tag{21}$$

where $r = R_t / R_b$, $R_t = 1.2249(A_d^{1/3} + A_e^{1/3})$ and $R_b = 1.43998 Z_d Z_e / Q$. The released energy $Q$ is evaluated using the mass tables [60-62] and the liquid-drop-model radius constant $r_0 = 1.2249$fm.

### 3.1.2 The Universal Decay Law (UDL)

Starting from the α-like (extension to the heavier cluster of α-decay theory) R-matrix theory and the microscopic mechanism of the charged-particle emission, a new universal decay law (UDL) for α-decay and cluster decay modes was introduced [4, 64] by Qi et al., The model was presented in an interesting way, which made it possible to represent, on the same plot with a single straight line, the logarithm of the half lives minus some quantity versus one of the two parameters ($\chi'$ and $\rho'$) that depend on the atomic and mass numbers of the daughter and emitted particles as well as the $Q$ value. UDL relates the half-life of monopole radioactive decay with the $Q$ values of the outgoing particles as well as the masses and charges of the nuclei involved in the decay and can be written in the logarithmic form as,

$$\log_{10}(T_{1/2}) = aZ_c Z_d \sqrt{\frac{A}{Q_c}} + b\sqrt{AZ_c Z_d (A_d^{1/3} + A_c^{1/3})} + c \tag{22}$$

$$= a\chi' + b\rho' + c \tag{23}$$

where the quantity $A = \dfrac{A_d A_c}{A_d + A_c}$ and the constants $a = 0.4314$, $b = -0.4087$ and $c = -25.7725$ are the coefficient sets of eq. (23), determined by fitting to experiments of both α and cluster decays [4]. The effects that induce the clusterization in the parent nucleus are included in the term $b\rho' + c$. As this relation holds for the monopole radioactive decays of all clusters, it is called the Universal Decay Law (UDL) [4].

### 3.1.3 Scaling law of Horoi et al.,

In order to determine the half lives of both the alpha and cluster decays, a new empirical formula for cluster decay was introduced by Horoi et al., [66] and is given by the equation,

$$\log_{10} T_{1/2} = (a_1 \mu^x + b_1)[(Z_1 Z_2)^y / \sqrt{Q} - 7] + (a_2 \mu^x + b_2) \tag{24}$$

where $\mu$ is the reduced mass. The six parameters are $a_1 = 9.1$, $b_1 = -10.2$, $a_2 = 7.39$, $b_2 = -23.2$, $x = 0.416$ and $y = 0.613$.

The behavior of the cluster half lives computed within CPPM, with the neutron number of the daughter nuclei can be clearly seen from the figures 1 and 2 which represents the plot for $\log_{10}(T_{1/2})$ vs. neutron number of the daughter nuclei, for the cluster emission of various clusters from $^{286-318}118$ superheavy nuclei. Fig. 1 represents the plot for the cluster emission of $^{4}$He, $^{8,10}$Be, $^{12,14,16,18}$C, $^{16,18,20,22,24}$O and $^{22,24}$Ne from $^{286-318}118$ superheavy nuclei and the plot for the

cluster emission of $^{26,28}$Ne, $^{26,\ 28,30,32,34}$Mg, $^{30,32,34,36,38}$Si and $^{38,40,42,44}$S has been given in figure 2. It can be seen clearly from figure 1, that the minima of the logarithmic half-lives for most of these cluster emission are found for the decay leading to a daughter with N = 184. For eg. In the case of $^4$He emission from $^{304}$118, the minima of the logarithmic half lives is found for the decay leading to $^{300}$116 (N = 184) and in the case of $^{10}$Be emission from $^{308}$118, the minima of the logarithmic half lives is found for the decay leading to $^{298}$114 (N = 184). A minimum in the decay half lives corresponds to the greater barrier penetrability, which in turn indicates the neutron/proton shell closure of the daughter nuclei. This indicate the role of neutron magicity N = 184 in cluster radioactivity. The present experimental upper and lower limits of half lives favourable for the cluster decay measurements, are $10^{30}$s and $10^{-6}$s respectively and have been represented as dotted line in these figures. As can be seen from the figures, most of the decays in figure 1 and a few decays in figure 2 are well within these experimental limits and are hence favourable for measurements.

The comparison of the predicted cluster decay half lives with that of the cluster half lives evaluated using various theoretical models, for the emission of various clusters from $^{286-318}$118 has been given in the tables 1-4. Only the most probable cluster emissions, most of those with $T_{1/2} < 10^{30}$s, are given here. The parent nuclei, emitted cluster and the daughter nuclei are given in column 1, 2 and 3 respectively and the energy released in the decay has been given in column 4. In column 5, the cluster decay half lives evaluated within CPPM have been arranged. The decay half lives evaluated using the Universal Decay Law (UDL) of Qi et al., the Universal (UNIV) curve of Poenaru et al., and the Scaling Law of Horoi et al., are given in columns 6, 7 and 8 respectively. The alpha decay half life of the experimentally synthesised isotope $^{294}$118 has been evaluated within CPPM and is included in table 3. On comparison with the experimental half life, $T_{1/2} = 0.69 \times 10^{-3}$s [30], it can be seen that our value ($T_{1/2} = 2.508 \times 10^{-3}$s) matches well with the experimental value. In the recent study [47] on the investigation of the alpha decay half lives of $^{271-310}$118 superheavy nuclei, we had shown that, on inclusion of the deformations of both the parent and daughter nuclei, the predicted alpha decay half life of $^{294}$118 ($T_{1/2} = 0.53 \times 10^{-3}$s) is in better agreement with the experimental value. A comparison of the cluster decay half lives evaluated within CPPM with the half lives evaluated using various theoretical models shows that, for most of the decays, the CPPM values matches well with the UDL values than that of the UNIV or the values obtained using the Scaling Law of Horoi. As most of the cluster decay half lives predicted through our study are much below the experimental limit ($T_{1/2} < 10^{30}$s), these decays could be treated as favourable for measurements and hence we hope these observations to serve as a guide for the future experiments.

In the cluster decay studies on heavy nuclei, it has been shown that the half life is minimum for the decays leading to the doubly magic daughter $^{208}$Pb (Z = 82, N = 126) or its neighboring nuclei. The present study on the cluster decay half lives of the superheavy nuclei gives a pronounced minima for the daughter with N = 184. This may be interpreted as a result of the strong shell effect of the assumed magic number of the neutrons and this reveal that neutron shell closure plays a decisive role in the cluster decays of superheavy nuclei.

## 4. Conclusions

Taking the interacting barrier as the sum of Coulomb and proximity potential (within CPPM), the feasibility for the emission of $^4$He, $^{8,10}$Be, $^{12,14,16,18}$C, $^{16,18,20,22,24}$O, $^{22,24,26,28}$Ne, $^{26,28,30,32,34}$Mg, $^{30,32,34,36,38}$Si and $^{38,40,42,44}$S, from the superheavy nuclei with Z = 118 within the range 270 ≤ A ≤ 318 has been investigated. The cluster decay half lives have also been calculated using the Universal formula for cluster decay (UNIV) of Poenaru et al., the Universal Decay Law (UDL) and the Scaling Law of Horoi et al.,. A comparison of our calculated alpha and cluster half lives with the values evaluated within these theoretical models show a similar trend. The experimental and the predicted half life of the experimentally synthesised superheavy isotope $^{294}$118 are also found to be in agreement with each other. The plots for $\log_{10}(T_{1/2})$ against the neutron number of the daughter in the corresponding decay reveals that, for most of the decays, the half life is minimum for the decay leading to a daughter with N = 184. The predictions on the cluster decay half lives of Z = 118, performed within CPPM, may be of great use for further experimental investigation on cluster decay in the superheavy region, as most of the predicted half lives are well within the present upper limit for measurements.

**Acknowledgements**

The author KPS would like to thank the University Grants Commission, Govt. of India for the financial support under Major Research Project. No.42-760/2013 (SR) dated 22-03-2013.

**References**

[1] H. Becquerel, Compt. Rend. 122, 420 (1896).
[2] R. Bonetti, L. Milazzo Colli and G. D. Sassi, J. Phys. G: Nucl. Phys. 3, 1111 (1977).
[3] Z. Ren, C. Xu and Z. Wang, Phys. Rev. C 70, 034304 (2004).

[4] C. Qi, F. R. Xu, R. J. Liotta and R. Wyss, Phys. Rev. Lett. 103, 072501 (2009).

[5] D. Ni and Z. Ren, Phys. Rev. C 82, 024311 (2010).

[6] D. N. Poenaru, R. A. Gherghescu and W. Greiner, Phys. Rev. Lett. 107, 062503 (2011).

[7] D. N. Poenaru, R. A. Gherghescu and W. Greiner, Phys. Rev. C 83, 014601 (2011).

[8] M. Mirea, A. Sandulescu and D. S. Delion, Proc. Rom. Acad. Series A, 12, 203 (2011).

[9] D. N. Poenaru, R. A. Gherghescu and W. Greiner, J. Phys. G: Nucl. Part. Phys. 39, 015105 (2012).

[10] A. V. Karpov, V. I. Zagrebaev, W. Greiner, L. F. Ruiz and Y. M. Palenzuela, Int. J. Mod. Phys. E 21, 1250013 (2012).

[11] P. B. Price, Nucl. Phys. A 502, 41c (1989).

[12] H. J. Rose and G. A. Jones, Nature 307, 245 (1984).

[13] A. Sandulescu, D. N.Poenaru and W. Greiner, Sov. J. Part. Nucl. 11, 528 (1980).

[14] D. V. Aleksandrov, A. F. Belyatskii, Yu. A. Glukhov, E. Yu. Nikolskii, B. G. Novatskii, A. A. Oglobin and D. N. Stepanov, Pis'ma Zh. Eksp. Teor. Fiz. 40, 152 (1984) [JETP Lett. 40, 909 (1984).

[15] A. Sandulescu, Yu. S. Zamiatin, J. A. Lebedev, B. F. Myasoedev, S. P. Tretyakova and D. Hasegan, Izv. Akad. Nauk SSSR Ser. Fiz. 41, 2104 (1985).

[16] S. W. Barwick, P. B. Price and J. D. Stevenson, Phys. Rev. C 31, 1984 (1985).

[17] R. Bonetti, E. Pioretto, H. C. Migliorino, A. Pasinetti, F. Barranco, E. Vigezzi and R. A. Broglia, Phys. Lett. B 241, 179 (1990).

[18] R. Bonetti and A. Guglielmetti, Rom. Rep. Phys. 59, 301 (2007).

[19] B. John, IANCAS Bulletin, 7, 238 (2008).

[20] A. Sandulescu and W. Greiner, J. Phys. G 3, L189 (1977).

[21] A. Sandulescu, H. J. Lustig, J. Hahn and W. Greiner, J. Phys. G 4, L279 (1978).

[22] A. Sandulescu, D. N. Poenaru, W.Greiner and J. H. Hamilton, Phys. Rev. Lett. 54, 490 (1985).

[23] A. Sandulescu, J. Phys. G 15, 529 (1989).

[24] A. Sandulescu and W. Greiner, Rep. Prog. Phys. 55, 1423 (1992).

[25] A. Sandulescu, R. K. Gupta, W. Scheid and W. Greiner, Phys. Lett. B 60, 225 (1976).

[26] R. K. Gupta, C. Parvulescu, A. Sandulescu and W. Greiner, Z. Phys. A 283, 217 (1977).

[27] Yu. Ts. Oganessian, A. G. Demin, A. S. Iljinov, S. P. Tretyakova, A. A. Pleve, Yu. E. Penionzhkevich, M. P. Ivanov and Yu. P. Tretyakova, Nucl. Phys. A 239, 157 (1975).

[28] S. Hofmann and G. Munzenberg, Rev. Mod. Phys. 72, 733 (2000).

[29] Yu. Ts. Oganessian, V. K. Utyonkov, Yu. V. Lobanov, F. Sh. Abdullin, A. N. Polyakov, R. N. Sagaidak, I. V. Shirokovsky, Yu. S. Tsyganov, A. A. Voinov, G. G. Gulbekian, S. L. Bogomolov, B. N. Gikal, A. N. Mezentsev, S. Iliev, V. G. Subbotin, A. M. Sukhov, K. Subotic, V. I. Zagrebaev, G. K. Vostokin, M. G. Itkis, K. J. Moody, J. B. Patin, D. A. Shaughnessy, M. A. Stoyer, N. J. Stoyer, P. A. Wilk, J. M. Kenneally, J. H. Landrum, J. F. Wild and R. W. Lougheed, Phys. Rev. **C 74**, 044602 (2006).

[30] Yu. Ts. Oganessian, F. Sh. Abdullin, C. Alexander, J. Binder, R. A. Boll, S. N. Dmitriev, J. Ezold, K. Felker, J. M. Gostic, R. K. Grzywacz, J. H. Hamilton, R. A. Henderson, M. G. Itkis, K. Miernik, D. Miller, K. J. Moody, A. N. Polyakov, A. V. Ramayya, J. B. Roberto, M. A. Ryabinin, K. P. Rykaczewski, R. N. Sagaidak, D. A. Shaughnessy, I. V. Shirokovsky, M. V. Shumeiko, M. A. Stoyer, N. J. Stoyer, V. G. Subbotin, A. M. Sukhov, Yu. S. Tsyganov, V. K. Utyonkov, A. A. Voinov and G. K. Vostokin, Phys. Rev. Lett. **109**, 162501 (2012).

[31] H. J. Fink, J. Maruhn, W. Scheid and W. Greiner, Zeitschrift fuer Physik 268, 321 (1974).

[32] M. Irionda, D. Jerrestam and R. J. Liotta, Nucl. Phys. A 454, 252 (1986).

[33] R. Blendowske, T. Fliessbach and H. Walliser, Nucl. Phys. A 464, 75 (1987).

[34] S. S. Malik and R. K. Gupta, Phys. Rev. C 39, 1992 (1989).

[35] S. Kumar and R. K. Gupta, Phys. Rev. C 55, 218 (1997).

[36] Y. J. Shi, W.J. Swiatecki, Phys. Rev. Lett. 54, 300 (1985).

[37] G. Shanmugam, B.Kamalaharan, Phys. Rev. C 38, 1377(1988).

[38] B. Buck, A.C. Merchant, J. Phys. G: Nucl. Part. Phys. 15, 615 (1989).

[39] K.P. Santhosh, Antony Joseph, Pramana J. Phys. 55, 375(2000).

[40] K.P. Santhosh, Antony Joseph, Ind. J. Pure App. Phys.42, 806 (2004).

[41] G. Royer, R. K. Gupta, V. Yu. Denisov, Nucl. Phys. A 632, 275 (1998).

[42] K. P. Santhosh, S. Sabina and G. J. Jayesh, Nucl. Phys. A 850, 34 (2011).

[43] K. P. Santhosh, B. Priyanka, G. J. Jayesh and S. Sabina, Phys. Rev. C 84, 024609 (2011).

[44] K. P. Santhosh, B. Priyanka and M. S. Unnikrishnan, Phys. Rev. C 85, 034604 (2012).

[45] K. P. Santhosh and B. Priyanka, J. Phys. G: Nucl. Part. Phys. 39, 085106 (2012).

[46] K. P. Santhosh and B. Priyanka, Phys. Rev. C 87, 064611 (2013).

[47] K. P. Santhosh and B. Priyanka, Phys. Rev. C 89, 064604 (2014).

[48] K. P. Santhosh and B. Priyanka, Nucl. Phys. A 929, 20 (2014).

[49] K. P. Santhosh, V. Bobby Jose, A. Joseph and K. M. Varier, Nucl. Phys. A 817, 35 (2009).

[50] K. P. Santhosh and V. Bobby Jose, Nucl. Phys. A 922, 191 (2014).

[51] M. A. Stoyer, Nature 442, 876 (2006).

[52] Yu. Ts. Oganessian, V. K. Utyonkov, Yu. V. Lobanov, F. Sh. Abdullin, A. N. Polyakov, I. V. Shirokovsky, Yu. S. Tsyganov, G. G. Gulbekian, S. L. Bogomolov, B. N. Gikal, A. N. Mezentsev, S. Iliev, V. G. Subbotin, A. M. Sukhov, O. V. Ivanov, G. V. Buklanov, K. Subotic, A. A. Voinov, M. G. Itkis, K. J. Moody, J. F. Wild, N. J. Stoyer, M. A. Stoyer, R. W. Lougheed and C. A. Laue, Eur. Phys. J. A 15, 201 (2002).

[53] Y. J. Shi and W. J. Swiatecki, Nucl. Phys. A 438, 450 (1985).

[54] J. Blocki, J. Randrup, W. J. Swiatecki and C. F. Tsang, Ann. Phys. NY 105, 427 (1977).

[55] I. Dutt and R. K. Puri, Phys. Rev. C 81, 064608 (2010).

[56] I. Dutt and R. K. Puri, Phys. Rev. C 81 064609 (2010).

[57] J. Blocki and W. J. Swiatecki, Ann. Phys. NY 132, 53 (1981).

[58] Y. J. Shi and W. J. Swiatecki, Nucl. Phys. A 464, 205 (1987).

[59] D. N. Poenaru, M. Ivascu and A. Sandulescu, W. Greiner, Phys. Rev. C 32, 572 (1985).

[60] M. Wang, G. Audi, A. H. Wapstra, F. G. Kondev, M. MacCormick, X. Xu and B. Pfeiffer, Chin. Phys. C 36, 1603 (2012).

[61] H. Koura, T. Tachibana, M. Uno and M. Yamada, Prog. Theor. Phys.113, 305 (2005).

[62] P. Moller, W. D. Myers, W. J. Swiatecki and J. Treiner, At. Data Nucl. Data Tables 39, 185 (1988).

[63] S. Athanassopoulos, E. Mavrommatis, K. A. Gernoth, J. W. Clark, arXiv:nucl-th/0509075v1, 2005.

[64] C. Qi, F.R. Xu, R.J. Liotta, R. Wyss, M.Y. Zhang, C. Asawatangtrakuldee and D. Hu, Phys. Rev. C 80, 044326 (2009).

[65] D. N. Poenaru, R. A. Gherghescu and W. Greiner, Phys. Rev. C 85, 034615 (2012).

[66] M. Horoi, A. Brown and A. Sandulescu, arXiv:nucl-th/9403008v1, 1994.

[67] D. N. Poenaru and W. Greiner, J. Phys. G, Nucl. Part. Phys. 17, S443 (1991).

[68] D. N. Poenaru and W. Greiner, Phys. Scr. 44, 427 (1991).

[69] D. N. Poenaru, I. H. Plonski and W. Greiner, Phys. Rev. C 74, 014312 (2006).

[70] D. N. Poenaru, I. H. Plonski, R. A. Gherghescu and W. Greiner, J. Phys. G, Nucl. Part. Phys. 32, 1223 (2006).

[71] R. Blendowske and H. Walliser, Phys. Rev. Lett. 61, 1930 (1988).

[72] R. Blendowske, T. Fliessbach, H. Walliser, in: Nuclear Decay Modes, Institute of Physics Publishing, Bristol, 1996, p. 337 (Chapter 7).

**Table 1**. Comparison of the of predicted cluster decay half-lives with that of the cluster half-lives evaluated using various theoretical models, for the emission of various clusters from $^{286,\,288}$118 SHN. The half-lives are calculated for zero angular momentum transfers.

| Parent Nuclei | Emitted Cluster | Daughter Nuclei | Q value (MeV) | $T_{1/2}$ (s) CPPM | UNIV | UDL | Horoi |
|---|---|---|---|---|---|---|---|
| $^{286}$118 | $^{4}$He | $^{282}$116 | 12.335 | $1.919 \times 10^{-4}$ | $2.148 \times 10^{-5}$ | $3.452 \times 10^{-5}$ | $8.588 \times 10^{-5}$ |
| | $^{8}$Be | $^{278}$114 | 24.678 | $6.791 \times 10^{13}$ | $3.160 \times 10^{10}$ | $3.875 \times 10^{13}$ | $7.002 \times 10^{11}$ |
| | $^{12}$C | $^{274}$112 | 44.350 | $3.477 \times 10^{16}$ | $7.199 \times 10^{11}$ | $2.272 \times 10^{15}$ | $9.617 \times 10^{13}$ |
| | $^{14}$C | $^{272}$112 | 42.260 | $1.992 \times 10^{21}$ | $5.611 \times 10^{17}$ | $4.872 \times 10^{21}$ | $2.108 \times 10^{20}$ |
| | $^{16}$O | $^{270}$110 | 64.617 | $4.345 \times 10^{19}$ | $2.598 \times 10^{13}$ | $3.528 \times 10^{16}$ | $3.415 \times 10^{16}$ |
| | $^{18}$O | $^{268}$110 | 61.693 | $4.404 \times 10^{24}$ | $1.251 \times 10^{19}$ | $6.138 \times 10^{22}$ | $2.725 \times 10^{22}$ |
| | $^{22}$Ne | $^{264}$108 | 83.021 | $2.158 \times 10^{25}$ | $4.257 \times 10^{18}$ | $7.969 \times 10^{20}$ | $1.085 \times 10^{23}$ |
| | $^{24}$Ne | $^{262}$108 | 80.242 | $4.159 \times 10^{29}$ | $3.325 \times 10^{23}$ | $1.671 \times 10^{26}$ | $1.018 \times 10^{28}$ |
| | $^{26}$Mg | $^{260}$106 | 104.227 | $1.815 \times 10^{26}$ | $3.687 \times 10^{18}$ | $8.131 \times 10^{18}$ | $1.449 \times 10^{24}$ |
| | $^{28}$Mg | $^{258}$106 | 104.339 | $1.712 \times 10^{25}$ | $1.066 \times 10^{20}$ | $3.827 \times 10^{19}$ | $4.188 \times 10^{25}$ |
| | $^{30}$Si | $^{256}$104 | 124.769 | $5.016 \times 10^{27}$ | $9.685 \times 10^{18}$ | $1.279 \times 10^{17}$ | $6.072 \times 10^{25}$ |
| | $^{32}$Si | $^{254}$104 | 125.438 | $2.746 \times 10^{25}$ | $7.269 \times 10^{19}$ | $7.122 \times 10^{16}$ | $3.929 \times 10^{26}$ |
| | $^{34}$Si | $^{252}$104 | 121.157 | $1.352 \times 10^{31}$ | $6.956 \times 10^{24}$ | $5.027 \times 10^{22}$ | $3.721 \times 10^{31}$ |
| $^{288}$118 | $^{4}$He | $^{284}$116 | 11.905 | $1.897 \times 10^{-3}$ | $1.606 \times 10^{-4}$ | $3.301 \times 10^{-4}$ | $6.073 \times 10^{-4}$ |
| | $^{8}$Be | $^{280}$114 | 23.628 | $1.687 \times 10^{16}$ | $3.832 \times 10^{12}$ | $9.001 \times 10^{15}$ | $9.030 \times 10^{13}$ |
| | $^{12}$C | $^{276}$112 | 44.390 | $2.289 \times 10^{16}$ | $5.280 \times 10^{11}$ | $1.587 \times 10^{15}$ | $8.601 \times 10^{13}$ |
| | $^{14}$C | $^{274}$112 | 41.510 | $5.277 \times 10^{22}$ | $7.674 \times 10^{18}$ | $1.141 \times 10^{23}$ | $3.704 \times 10^{21}$ |
| | $^{18}$O | $^{270}$110 | 60.843 | $9.719 \times 10^{25}$ | $1.228 \times 10^{20}$ | $1.120 \times 10^{24}$ | $3.444 \times 10^{23}$ |
| | $^{22}$Ne | $^{266}$108 | 81.625 | $2.292 \times 10^{27}$ | $9.560 \times 10^{19}$ | $5.207 \times 10^{22}$ | $3.257 \times 10^{24}$ |
| | $^{24}$Ne | $^{264}$108 | 81.128 | $7.700 \times 10^{27}$ | $2.459 \times 10^{22}$ | $5.443 \times 10^{24}$ | $1.045 \times 10^{27}$ |
| | $^{28}$Mg | $^{260}$106 | 103.211 | $5.721 \times 10^{26}$ | $8.514 \times 10^{20}$ | $7.514 \times 10^{20}$ | $4.935 \times 10^{26}$ |
| | $^{32}$Si | $^{256}$104 | 124.595 | $2.762 \times 10^{26}$ | $2.459 \times 10^{20}$ | $4.549 \times 10^{17}$ | $2.032 \times 10^{27}$ |
| | $^{34}$Si | $^{254}$104 | 121.497 | $1.753 \times 10^{30}$ | $2.485 \times 10^{24}$ | $1.139 \times 10^{22}$ | $1.969 \times 10^{31}$ |

**Table 2**. Comparison of the of predicted cluster decay half-lives with that of the cluster half-lives evaluated using various theoretical models, for the emission of various clusters from $^{290, 292}118$ SHN. The half-lives are calculated for zero angular momentum transfers.

| Parent Nuclei | Emitted Cluster | Daughter Nuclei | Q value (MeV) | $T_{1/2}$ (s) | | | |
|---|---|---|---|---|---|---|---|
| | | | | CPPM | UNIV | UDL | Horoi |
| $^{290}118$ | $^{4}$He | $^{286}116$ | 11.645 | $7.817 \times 10^{-3}$ | $5.601 \times 10^{-4}$ | $1.337 \times 10^{-3}$ | $2.091 \times 10^{-3}$ |
| | $^{8}$Be | $^{282}114$ | 22.728 | $2.511 \times 10^{18}$ | $3.120 \times 10^{14}$ | $1.279 \times 10^{18}$ | $7.600 \times 10^{15}$ |
| | $^{12}$C | $^{278}112$ | 42.720 | $1.744 \times 10^{19}$ | $1.011 \times 10^{14}$ | $9.304 \times 10^{17}$ | $1.944 \times 10^{16}$ |
| | $^{14}$C | $^{276}112$ | 42.260 | $1.170 \times 10^{21}$ | $3.775 \times 10^{17}$ | $3.160 \times 10^{21}$ | $2.191 \times 10^{20}$ |
| | $^{18}$O | $^{272}110$ | 60.393 | $4.358 \times 10^{26}$ | $3.778 \times 10^{20}$ | $4.708 \times 10^{24}$ | $1.365 \times 10^{24}$ |
| | $^{22}$Ne | $^{268}108$ | 80.825 | $2.872 \times 10^{28}$ | $5.342 \times 10^{20}$ | $5.226 \times 10^{23}$ | $2.415 \times 10^{25}$ |
| | $^{24}$Ne | $^{266}108$ | 80.442 | $7.219 \times 10^{28}$ | $1.113 \times 10^{23}$ | $4.144 \times 10^{25}$ | $6.484 \times 10^{27}$ |
| | $^{28}$Mg | $^{262}106$ | 102.279 | $9.792 \times 10^{27}$ | $4.693 \times 10^{21}$ | $8.576 \times 10^{21}$ | $3.938 \times 10^{27}$ |
| | $^{32}$Si | $^{258}104$ | 123.368 | $1.123 \times 10^{28}$ | $1.758 \times 10^{21}$ | $8.787 \times 10^{18}$ | $2.241 \times 10^{28}$ |
| | $^{34}$Si | $^{256}104$ | 121.364 | $1.379 \times 10^{30}$ | $2.309 \times 10^{24}$ | $1.049 \times 10^{22}$ | $2.732 \times 10^{31}$ |
| $^{292}118$ | $^{4}$He | $^{288}116$ | 11.465 | $2.096 \times 10^{-2}$ | $1.339 \times 10^{-3}$ | $3.545 \times 10^{-3}$ | $5.049 \times 10^{-3}$ |
| | $^{8}$Be | $^{284}114$ | 22.208 | $4.850 \times 10^{19}$ | $4.302 \times 10^{15}$ | $2.435 \times 10^{19}$ | $1.118 \times 10^{17}$ |
| | $^{12}$C | $^{280}112$ | 41.220 | $9.058 \times 10^{21}$ | $1.566 \times 10^{16}$ | $3.925 \times 10^{20}$ | $3.352 \times 10^{18}$ |
| | $^{14}$C | $^{278}112$ | 40.990 | $3.868 \times 10^{23}$ | $3.833 \times 10^{19}$ | $8.116 \times 10^{23}$ | $2.906 \times 10^{22}$ |
| | $^{18}$O | $^{274}110$ | 58.523 | $8.300 \times 10^{29}$ | $1.064 \times 10^{23}$ | $5.442 \times 10^{27}$ | $4.552 \times 10^{26}$ |
| | $^{24}$Ne | $^{268}108$ | 80.042 | $2.243 \times 10^{29}$ | $2.421 \times 10^{23}$ | $1.191 \times 10^{26}$ | $1.930 \times 10^{28}$ |
| | $^{28}$Mg | $^{264}106$ | 101.159 | $3.537 \times 10^{29}$ | $4.095 \times 10^{22}$ | $1.837 \times 10^{23}$ | $4.922 \times 10^{28}$ |
| | $^{32}$Si | $^{260}104$ | 121.848 | $1.359 \times 10^{30}$ | $2.343 \times 10^{22}$ | $4.157 \times 10^{20}$ | $4.559 \times 10^{29}$ |
| | $^{34}$Si | $^{258}104$ | 120.537 | $1.535 \times 10^{31}$ | $8.803 \times 10^{24}$ | $7.675 \times 10^{22}$ | $1.575 \times 10^{32}$ |

**Table 3**. Comparison of the of predicted cluster decay half-lives with that of the cluster half-lives evaluated using various theoretical models, for the emission of various clusters from $^{294-304}118$ SHN. The half-lives are calculated for zero angular momentum transfers.

| Parent Nuclei | Emitted Cluster | Daughter Nuclei | Q value (MeV) | $T_{1/2}$ (s) CPPM | UNIV | UDL | Horoi |
|---|---|---|---|---|---|---|---|
| $^{294}118$ | $^{4}$He | $^{290}116$ | 11.815 | 2.508x10$^{-3}$ | 2.022x10$^{-4}$ | 4.381x10$^{-4}$ | 9.347x10$^{-4}$ |
|  | $^{8}$Be | $^{286}114$ | 22.718 | 1.937x10$^{18}$ | 2.490x10$^{14}$ | 1.029x10$^{18}$ | 8.171x10$^{15}$ |
|  | $^{12}$C | $^{282}112$ | 40.450 | 2.271x10$^{23}$ | 2.181x10$^{17}$ | 9.074x10$^{21}$ | 5.309x10$^{19}$ |
|  | $^{14}$C | $^{280}112$ | 40.550 | 2.608x10$^{24}$ | 1.779x10$^{20}$ | 5.146x10$^{24}$ | 1.689x10$^{23}$ |
|  | $^{24}$Ne | $^{270}108$ | 80.132 | 1.003x10$^{29}$ | 1.442x10$^{23}$ | 6.154x10$^{25}$ | 1.578x10$^{28}$ |
|  | $^{28}$Mg | $^{266}106$ | 100.669 | 1.269x10$^{30}$ | 9.060x10$^{22}$ | 5.709x10$^{23}$ | 1.543x10$^{29}$ |
|  | $^{32}$Si | $^{262}104$ | 120.958 | 1.749x10$^{31}$ | 9.683x10$^{22}$ | 3.422x10$^{21}$ | 2.791x10$^{30}$ |
| $^{296}118$ | $^{4}$He | $^{292}116$ | 10.125 | 1.157x10$^{2}$ | 3.118x10$^{0}$ | 1.781x10$^{1}$ | 7.230x10$^{0}$ |
|  | $^{8}$Be | $^{288}114$ | 20.808 | 2.764x10$^{23}$ | 9.796x10$^{18}$ | 1.323x10$^{23}$ | 2.530x10$^{20}$ |
|  | $^{12}$C | $^{284}112$ | 38.240 | 6.232x10$^{27}$ | 1.040x10$^{21}$ | 1.896x10$^{26}$ | 2.262x10$^{23}$ |
|  | $^{14}$C | $^{282}112$ | 38.630 | 3.677x10$^{28}$ | 4.246x10$^{23}$ | 5.027x10$^{28}$ | 4.840x10$^{26}$ |
| $^{298}118$ | $^{4}$He | $^{294}116$ | 11.115 | 1.424x10$^{-1}$ | 7.246x10$^{-3}$ | 2.355x10$^{-2}$ | 2.989x10$^{-2}$ |
|  | $^{14}$C | $^{284}112$ | 37.890 | 1.516x10$^{30}$ | 9.116x10$^{24}$ | 1.846x10$^{30}$ | 1.239x10$^{28}$ |
| $^{300}118$ | $^{4}$He | $^{296}116$ | 11.035 | 2.192x10$^{-1}$ | 1.057x10$^{-2}$ | 3.601x10$^{-2}$ | 4.547x10$^{-2}$ |
|  | $^{8}$Be | $^{292}114$ | 21.448 | 3.069x10$^{21}$ | 1.714x10$^{17}$ | 1.568x10$^{21}$ | 6.970x10$^{18}$ |
|  | $^{14}$C | $^{286}112$ | 37.780 | 2.189x10$^{30}$ | 1.226x10$^{25}$ | 2.669x10$^{30}$ | 2.061x10$^{28}$ |
| $^{302}118$ | $^{4}$He | $^{298}116$ | 10.915 | 4.389x10$^{-1}$ | 1.950x10$^{-2}$ | 7.135x10$^{-2}$ | 8.590x10$^{-2}$ |
|  | $^{8}$Be | $^{294}114$ | 21.088 | 2.757x10$^{22}$ | 1.215x10$^{18}$ | 1.395x10$^{22}$ | 5.282x10$^{19}$ |
|  | $^{14}$C | $^{288}112$ | 37.920 | 8.194x10$^{29}$ | 5.359x10$^{24}$ | 1.043x10$^{30}$ | 1.133x10$^{28}$ |
| $^{304}118$ | $^{4}$He | $^{300}116$ | 12.435 | 5.793x10$^{-5}$ | 7.025x10$^{-6}$ | 1.057x10$^{-5}$ | 5.664x10$^{-5}$ |
|  | $^{8}$Be | $^{296}114$ | 22.478 | 3.868x10$^{18}$ | 4.448x10$^{14}$ | 2.143x10$^{18}$ | 2.933x10$^{16}$ |
|  | $^{10}$Be | $^{294}114$ | 19.303 | 2.538x10$^{31}$ | 1.472x10$^{28}$ | 8.722x10$^{32}$ | 3.235x10$^{31}$ |
|  | $^{14}$C | $^{290}112$ | 39.810 | 3.343x10$^{25}$ | 1.331x10$^{21}$ | 6.240x10$^{25}$ | 3.720x10$^{24}$ |

Table 4. Comparison of the of predicted cluster decay half-lives with that of the cluster half-lives evaluated using various theoretical models, for the emission of various clusters from $^{306-318}118$ SHN. The half-lives are calculated for zero angular momentum transfers.

| Parent Nuclei | Emitted Cluster | Daughter Nuclei | Q value (MeV) | $T_{1/2}$ (s) CPPM | UNIV | UDL | Horoi |
|---|---|---|---|---|---|---|---|
| $^{306}118$ | $^{4}$He | $^{302}116$ | 11.895 | $1.046 \times 10^{-3}$ | $8.778 \times 10^{-5}$ | $1.815 \times 10^{-4}$ | $6.527 \times 10^{-4}$ |
| | $^{10}$Be | $^{296}114$ | 20.673 | $3.846 \times 10^{26}$ | $6.796 \times 10^{23}$ | $1.530 \times 10^{28}$ | $1.590 \times 10^{27}$ |
| | $^{14}$C | $^{292}112$ | 41.080 | $4.989 \times 10^{22}$ | $6.939 \times 10^{18}$ | $1.228 \times 10^{23}$ | $2.333 \times 10^{22}$ |
| $^{308}118$ | $^{4}$He | $^{304}116$ | 11.295 | $3.332 \times 10^{-2}$ | $1.867 \times 10^{-3}$ | $5.492 \times 10^{-3}$ | $1.209 \times 10^{-2}$ |
| | $^{10}$Be | $^{298}114$ | 21.793 | $9.191 \times 10^{22}$ | $4.108 \times 10^{20}$ | $4.153 \times 10^{24}$ | $9.685 \times 10^{23}$ |
| | $^{14}$C | $^{294}112$ | 41.920 | $7.264 \times 10^{20}$ | $2.356 \times 10^{17}$ | $2.160 \times 10^{21}$ | $9.301 \times 10^{20}$ |
| $^{310}118$ | $^{4}$He | $^{306}116$ | 10.275 | $2.463 \times 10^{1}$ | $7.060 \times 10^{-1}$ | $3.774 \times 10^{0}$ | $3.066 \times 10^{0}$ |
| | $^{10}$Be | $^{300}114$ | 20.203 | $1.109 \times 10^{28}$ | $1.300 \times 10^{25}$ | $4.137 \times 10^{29}$ | $4.432 \times 10^{28}$ |
| | $^{14}$C | $^{296}112$ | 42.070 | $2.927 \times 10^{20}$ | $1.118 \times 10^{17}$ | $8.972 \times 10^{20}$ | $5.359 \times 10^{20}$ |
| | $^{20}$O | $^{290}110$ | 56.934 | $2.182 \times 10^{32}$ | $6.978 \times 10^{26}$ | $4.355 \times 10^{31}$ | $6.311 \times 10^{31}$ |
| $^{312}118$ | $^{4}$He | $^{308}116$ | 9.275 | $4.432 \times 10^{4}$ | $6.680 \times 10^{2}$ | $6.392 \times 10^{3}$ | $1.656 \times 10^{3}$ |
| | $^{14}$C | $^{298}112$ | 39.880 | $1.035 \times 10^{25}$ | $4.647 \times 10^{20}$ | $1.956 \times 10^{25}$ | $3.011 \times 10^{24}$ |
| | $^{16}$C | $^{296}112$ | 36.976 | $1.438 \times 10^{33}$ | $2.678 \times 10^{29}$ | $7.546 \times 10^{34}$ | $7.440 \times 10^{33}$ |
| | $^{22}$O | $^{290}110$ | 57.030 | $2.308 \times 10^{32}$ | $8.319 \times 10^{28}$ | $1.869 \times 10^{33}$ | $1.892 \times 10^{34}$ |
| $^{314}118$ | $^{4}$He | $^{310}116$ | 9.035 | $3.061 \times 10^{5}$ | $3.931 \times 10^{3}$ | $4.338 \times 10^{4}$ | $8.767 \times 10^{3}$ |
| | $^{14}$C | $^{300}112$ | 37.920 | $2.400 \times 10^{29}$ | $1.671 \times 10^{24}$ | $3.073 \times 10^{29}$ | $1.282 \times 10^{28}$ |
| | $^{22}$O | $^{292}110$ | 57.240 | $6.239 \times 10^{31}$ | $3.001 \times 10^{28}$ | $5.405 \times 10^{32}$ | $8.995 \times 10^{33}$ |
| $^{316}118$ | $^{4}$He | $^{312}116$ | 8.815 | $1.922 \times 10^{6}$ | $2.129 \times 10^{4}$ | $2.680 \times 10^{5}$ | $4.289 \times 10^{4}$ |
| | $^{22}$O | $^{294}110$ | 57.300 | $3.577 \times 10^{31}$ | $1.886 \times 10^{28}$ | $3.111 \times 10^{32}$ | $7.450 \times 10^{33}$ |
| $^{318}118$ | $^{4}$He | $^{314}116$ | 8.575 | $1.559 \times 10^{7}$ | $1.470 \times 10^{5}$ | $2.136 \times 10^{6}$ | $2.597 \times 10^{5}$ |

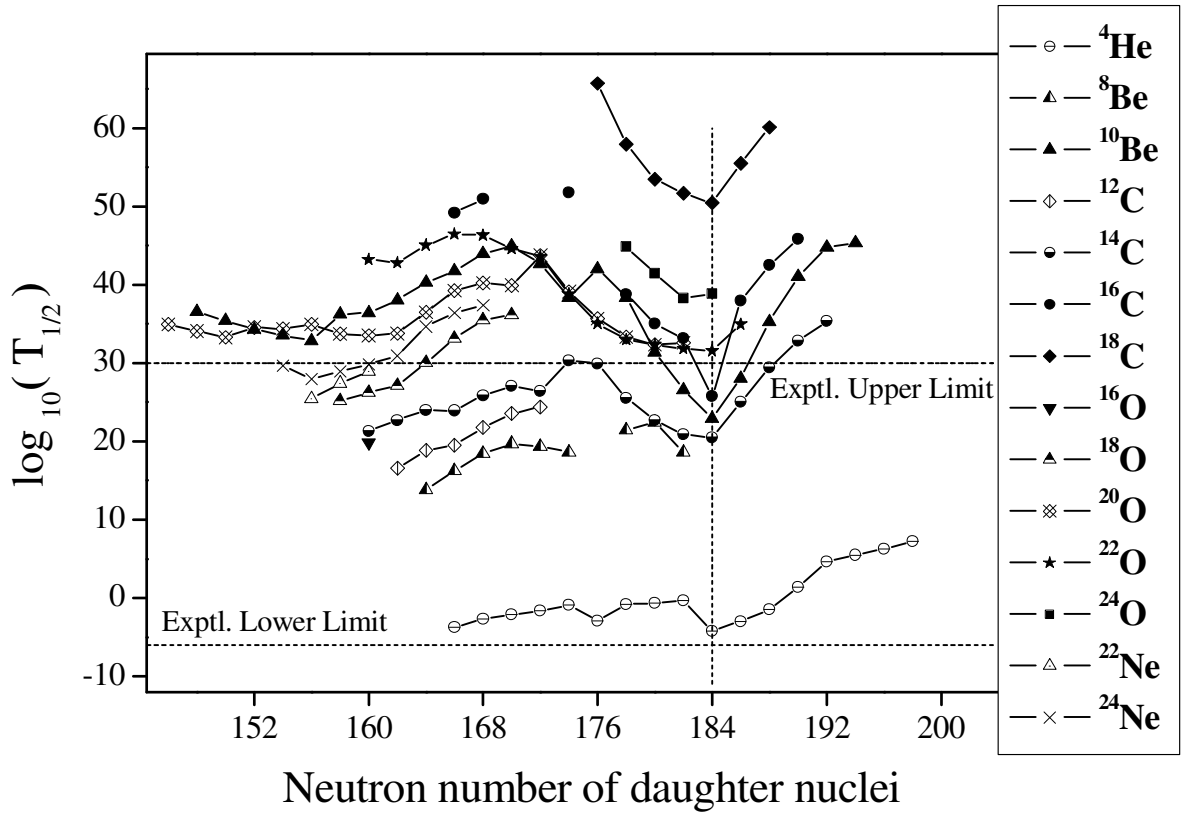

**Figure 1.** The computed $\log_{10}(T_{1/2})$ values vs. neutron number of daughter nuclei for the emission of various clusters from $^{286\text{-}318}$118 SHN. $T_{1/2}$ is in seconds.

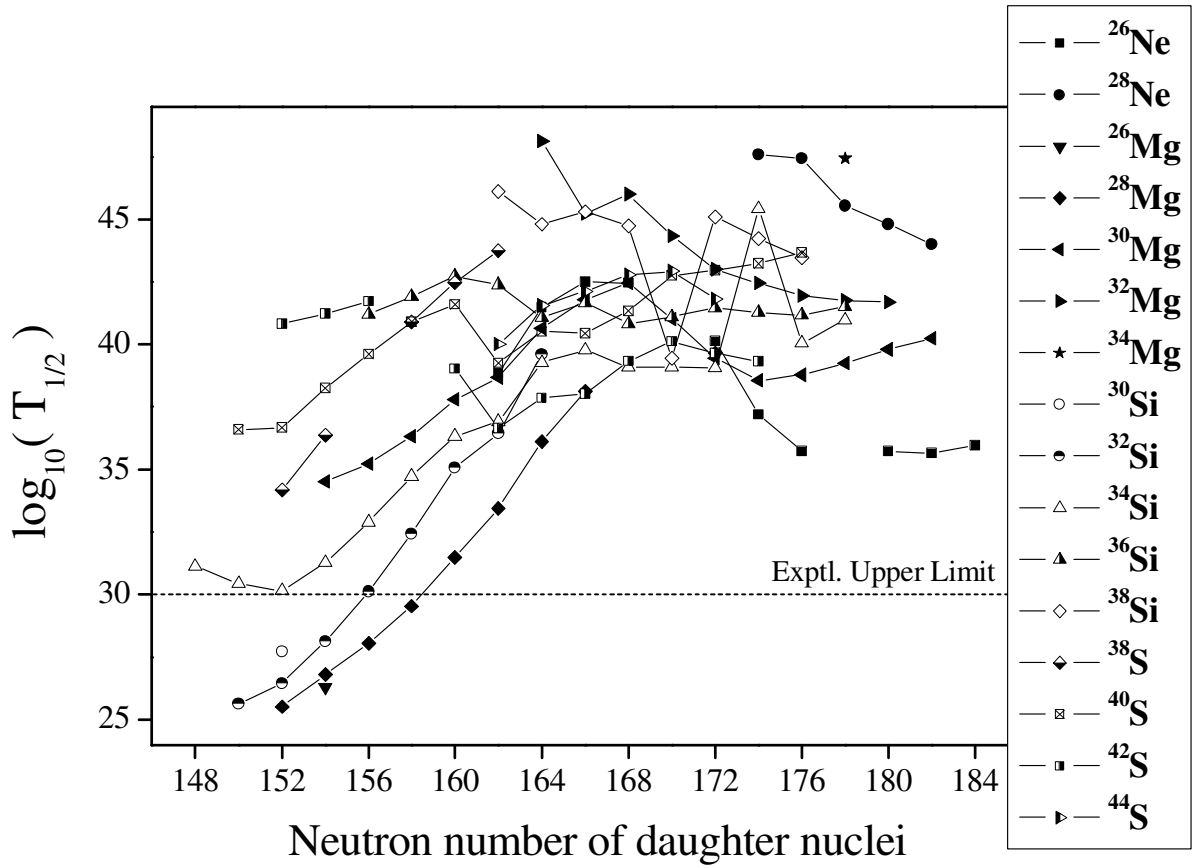

**Figure 2.** The computed $\log_{10}(T_{1/2})$ values vs. neutron number of daughter nuclei for the emission of various clusters from $^{286-318}$118 SHN. $T_{1/2}$ is in seconds.